\documentstyle[aps,prl,twocolumn,epsf]{revtex}

\begin{document}

\draft

\twocolumn[\hsize\textwidth\columnwidth\hsize\csname
@twocolumnfalse\endcsname

\tightenlines

\title{The Phase Diagram of Star Polymer Solutions}

\author{M.~Watzlawek,$^{1,}$\cite{author}
        C.~N.~Likos,$^{2}$ 
        and H.~L{\"o}wen$^{1,2}$
       }

\address{$^{1}$Institut f{\"u}r Theoretische Physik II,
               Heinrich-Heine-Universit{\"a}t, Universit{\"a}tsstra{\ss}e 1,
               D-40225 D{\"u}sseldorf, Germany \\
         $^{2}$IFF Theorie II,                         
               Forschungszentrum J{\"u}lich GmbH,       
               D-52425 J{\"u}lich, Germany
        \\~\\
        }

\date{{\em Phys. Rev. Lett.} {\bf 82}, 5289 (1999)}

\maketitle

\begin{abstract}
   The phase diagram of star polymer solutions in a good solvent
   is obtained over a wide range of densities and arm numbers by  
   Monte Carlo simulations. The effective interaction between the stars
   is modeled by an ultrasoft pair potential which is logarithmic in 
   the core-core distance. Among the stable phases are a fluid as well as 
   body-centered cubic, face-centered cubic, body-centered
   orthogonal, and diamond crystals.
   In a limited range of arm numbers, reentrant melting and 
   reentrant freezing transitions occur for increasing density. 
\end{abstract}

\vskip2pc]

\renewcommand{\thepage}{\hskip 8.9cm \arabic{page} \hfill Typeset
                        using REV\TeX }

\narrowtext

A major challenge in statistical physics is to understand and predict the
macroscopic phase behavior from a microscopic many-body theory for a 
given interaction between the particles \cite{Baus:book}. 
For a simple classical fluid \cite{Hansen:McDonald}, 
this interaction is specified in terms of a radially symmetric pair 
potential $V(r)$ where $r$ is the particle se\-pa\-ra\-tion.
Significant progress has been made during the last decades in predicting the
thermodynamically stable phases for simple intermolecular pair potentials, 
such as for Lennard-Jones systems, plasmas or hard spheres, 
using computer simulations \cite{Baus:book} and
density functional theory \cite{Loewen:report}. An important realization
of classical many-body systems are suspensions of colloidal
particles dispersed in a fluid medium. A striking advantage
of such colloidal samples over molecular ones is that 
their effective pair interaction is eminently tunable through
experimental control of particle and solvent properties 
\cite{Pusey:LH}. This brings about more
extreme pair interactions, leading to novel
phase transformations. For instance, if the colloidal particles are 
sterically stabilized against coagulation,
the `softness'  of the interparticle repulsion is governed
by the length of the polymer chains grafted onto the colloidal surface, 
their surface grafting density and solvent quality. 
Computer simulations and theory have revealed that 
a fluid freezes into a body-centered-cubic (bcc) crystal for soft 
long-ranged repulsions and into a
face-centered-cubic (fcc) one for strong short-ranged repulsions
\cite{Robbins:Kremer}. 
This was confirmed in experiments on sterically stabilized colloidal 
particles \cite{McConnell}.
A similar behavior occurs for charge-stabilized suspensions where 
the softness of $V(r)$
is now controlled by the concentration of added salt \cite{Sirota:89}.
Less common effects were
observed for potentials involving an attractive part aside from a 
repulsive core. In reducing the range of the attraction, a 
vanishing liquid phase has been observed \cite{Ilett:95} and  
an isostructural solid-solid transition was predicted 
\cite{Bolhuis:Frenkel}. More complicated pair potentials  
can even lead to stable quasicrystalline phases  
and a quadruple point in the phase diagram \cite{Denton}.

The aim of this letter is to study the phase diagram of an 
{\it ultrasoft} repulsive pair potential $V(r)$ which is 
lo\-ga\-rith\-mic in $r$ inside a core of diameter $\sigma$
and vanishes exponentially in $r$ outside the core. The motivation to do this 
is twofold: first, such a potential is a good model for the effective
interaction between star polymers in a good solvent 
\cite{Witten:Pincus,Likos:stars}, which can be regarded as 
sterically stabilized particles where
the size of the particles is much smaller than the length of 
the grafted polymer chains \cite{Grest:review}. These stars are 
characterized by their arm number (or functionality) $f$, i.e., the
number of polymer chains tethered to the central particle, 
and their corona diameter $\sigma$ which measures the spatial 
extension of the monomer density around a single star center.
Second, more fundamentally, phase transitions for such soft potentials
are expected to be 
rather different from that for stronger repulsions. From a study of 
the pure logarithmic potential in two spatial dimensions
\cite{Hansen:Levesque}, it is known that one needs a critical prefactor
to freeze the system, which is quite different from, e.g., inverse-power 
potentials. Furthermore, the potential crossover at $r=\sigma$ is expected 
to influence drastically the freezing transition, if the number density  
$\rho$ of the stars is near the overlap concentration, 
$\rho^{*}\approx 1/\sigma^3$.

We obtain the full phase diagram of star polymer solutions
by Monte Carlo simulation and theory. As a result, among the stable phases are 
a fluid as well as  bcc, fcc, body-centered-orthogonal (bco), 
and diamond crystals. We emphasize that the stability of a 
bco crystal with anisotropic rectangular elementary cell
and a diamond structure was never obtained before for a 
{\it radially symmetric} 
pair potential. In fact, there is a widespread belief
in the literature that anisotropic or three-body forces are solely 
responsible for a stable diamond lattice \cite{noDiamond}. 
We show that both the crossover at $r=\sigma$ and
the ultrasoftness of the core are crucial for the stability 
of the bco and the diamond phase.
Moreover, we get reentrant melting for $34 {<\atop\sim} f {<\atop\sim} 60$, 
and reentrant freezing for $44 {<\atop\sim} f {<\atop\sim} 60$
as $\rho$ is increasing. 
Some features of the presented phase diagram have already been observed
in a syatem of copolymer micelles exhibiting a very similar interaction to 
star polymers \cite{McConnell,McConnell2}.

With $k_BT$ denoting the thermal energy, 
our effective pair potential between two star centers is a combination 
of a logarithm inside the core of size $\sigma$ and a Yukawa-potential 
outside the core \cite{Likos:stars}:
\begin{equation} 
   \label{potential}
   V(r) = \frac{5}{18} k_BT f^{3/2} 
          \cases{
                   -\ln(\frac{r}{\sigma})+\frac{1}{1+\sqrt{f}/2}
                   & ($r\leq\sigma$);
           \cr 
                   \frac{\sigma}{1+\sqrt{f}/2}
                   \frac{\exp(-\sqrt{f}(r-\sigma)/2\sigma)}{r}  
                   & ($r>\sigma$),
           \cr}
\end{equation}
such that both the potential and its first derivative 
(or, equivalently, the force)
are continuous at $r=\sigma$ \cite{Likos:stars}. The decay length of the 
exponential is given by the largest blob diameter within 
the Daoud-Cotton theory for single star polymers \cite{Daoud:Cotton}. 
Experimental support for this potential comes from neutron scattering 
data on the structural ordering of 18-arm stars in the fluid phase 
\cite{Likos:stars} and shear moduli measurements in the crystalline 
phase of micelles \cite{Buitenhuis:Foerster}.
Furthermore, microscopic simulations of two star 
polymers have shown that this potential 
provides an excellent description of
the effective star interaction for a broad range of arm 
numbers \cite{Jusufi}. We note that $V(r)$ becomes the 
hard sphere potential for $f\to\infty$. 

Due to the purely entropic origin of the interstar repulsion, 
the strength of the pair potential (\ref{potential}) scales linearly 
with $k_{B}T$, causing the temperature to be
an irrelevant thermodynamic quantity.
Therefore, for the calculation of the phase diagram, only the packing
fraction of the stars, $\eta=\pi/6\rho\sigma^{3}$, and the arm number $f$
matter, the latter playing the role of an ``effective inverse temperature.''
We use computer simulations to access the phase diagram. 
The free energies of the
fluid phase and several possible solid phases are calculated by
thermodynamic integration via Monte Carlo simulations \cite{Frenkel:Smit}.
The free energy of the fluid phase, $F_{fl}$, is obtained either by the
well-known ``pressure- or density-route'' 
\cite{Hansen:McDonald,Frenkel:Smit}, or, alternatively, by the so-called
``$f$-route''. The pressure-route relates
the free energy for nonvanishing 
$\eta$ to that at zero packing fraction, keeping $f$
fixed. In the $f$-route, $f$
is used as an artificial thermodynamic variable, now keeping 
$\eta$ fixed. The free energy of star polymers with a certain 
arm number $f$ is then obtained by the following integration:
\begin{equation}
   \label{td.integration}
   F_{fl}=\int_{0}^{f}~df'
          \left\langle
             \frac{\partial U}{\partial f'}        
          \right\rangle_{f'}.
\end{equation}
Here, $U=\sum_{i<j}V(|{\bf r}_{i}-{\bf r}_{j}|)$ is the total
potential energy function which depends on $f$ since $V(r)$ depends on $f$ 
parametrically. $\langle ...\rangle_{f'}$ 
denotes the canonical ensemble average for a system with 
fixed arm number $f'$. Therefore, in order to
carry out the $f$-route integration, a series of
simulations at fixed $\eta$ but for increasing $f'$ is performed
to calculate the integrand of Eq.\ (\ref{td.integration}).

\begin{figure}
   \epsfxsize=6.5cm
   \epsfysize=5cm
   ~\hfill\epsfbox{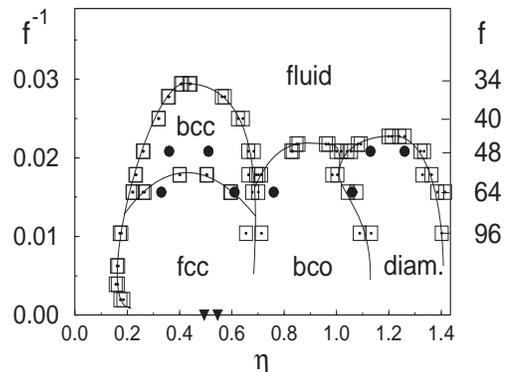}\hfill~
   \caption{The phase diagram of star polymer solutions for different arm 
            numbers $f$ versus packing fraction $\eta$. 
            The squares and the circles
            indicate the phase boundaries as obtained from computer
            simulations and theory, respectively; lines are only 
            guides to the eye. The statistical error of the simulations
            is of the order of the symbol size. The triangles indicate 
            the freezing and melting point of hard spheres.
           }
   \label{diagram.plot}
\end{figure}

We use the Frenkel-Ladd method for continuous potentials
to obtain the free energy of the solid phases
\cite{Frenkel:Smit,Frenkel:helium}. For these Monte Carlo calculations, 
suitable candidate crystal structures have 
to be chosen. Our method to get information about the possible stable
structures for fixed $f$ and $\eta$ consists of two steps: 
first, we calculate lattice sums for a wide class of crystals, 
including the `usual' structures
with cubic elementary cells (fcc, bcc,  hcp, and simple cubic)
and several `unusual' structures. These unusual structures are
the hexagonal lattice, the diamond lattice, representations of 
quasicrystalline structures (see, e.g., Ref.\ \cite{Denton}), and 
generalizations of the usual structures, which were obtained by 
stretching the elementary cell lengths (denoted as $a$, $b$, and $c$) 
of these structures by arbitrary factors,
then using the two independent ratios $b/a$ and $c/a$ as minimization 
parameters of the lattice sum. Second, we calculate the global bond order 
parameters \cite{bondorder:refs} of the equilibrated structures, 
which were spontaneously formed in a first set of simulations, 
always starting from a purely random configuration. 
Crystal structures whose bond order 
parameters are in agreement with these measured parameters, and which have 
reasonably small values of the lattice sum, are
then chosen as candidate structures for the 
free energy calculations. This procedure was performed for a wide range 
of arm numbers, $18\le f\le512$, and
packing fractions $0\le\eta\le1.5$. Finally, the obtained free energy data 
at fixed $f$ were used to explore the phase boundaries via the common
double tangent construction. The resulting phase diagram is displayed in
Fig.\ \ref{diagram.plot}. 
In the explored range of $f$ and $\eta$, four different
stable crystal structures are found besides a fluid phase.
For $f<f_{c}\simeq 34$, the fluid phase is stable for 
{\em all} densities, which is in agreement with results obtained from a 
effective hard sphere mapping procedure \cite{Watzlawek:98} and from
scaling theory \cite{Witten:Pincus}. We remark that Witten {\it et al.}
\cite{Witten:Pincus} only estimated $f_{c}$ within one order of magnitude
to be around $f\simeq 100$. For $f\ge f_{c}$, at least one 
stable crystal phase is found. We focus first on the crystal phases 
at $0.2{<\atop\sim}\eta{<\atop\sim}0.7$: 
for $f_{c}<f{<\atop\sim} 54$, a bcc phase is found, 
whereas for $f{>\atop\sim}70$, only the
fcc structure turns out to be stable. At intermediate
$f$ ($54{<\atop\sim}f{<\atop\sim}70$), bcc-fcc phase transitions occur. For   
$0.2{<\atop\sim}\eta{<\atop\sim}0.7$, the mean interparticle 
distance $\bar{r}=\rho^{-1/3}$ is larger than $\sigma$, leaving 
only the exponential part of $V(r)$ to be relevant for the phase 
behavior. Therefore, the observance of a fcc phase for large $f$, 
corresponding to a short-range, strongly screened potential, and a 
bcc phase for small $f$, corresponding to a long-range, less 
screened potential, is analogous to the phase behavior found for charged 
colloids \cite{Robbins:Kremer,Sirota:89}.
In Fig.\ \ref{diagram.plot}, the freezing and melting points for
hard spheres, corresponding to $f\to\infty$, are shown 
as well, denoted by black triangles.
We emphasize that even star polymers with very high arm numbers 
freeze at considerably smaller $\eta$ than 
hard spheres. In fact, our simulations show that a `hard-sphere like'
structure is only found for extremely high arm numbers
$f{>\atop\sim}10000$. Thus  
the change in the phase boundary cannot be shown
on the scale of the figure.
 
Let us now consider the phase behavior for $\eta\simeq0.7$, where $\bar{r}$ is
in the order of $\sigma$ and the logarithmic part of
$V(r)$ becomes relevant. From our calculations, a reentrant melting transition,
i.e., a transition from a solid to a liquid phase with increasing $\eta$, 
is found for $34<f{<\atop\sim}60$. We note that this reentrant melting
was already predicted qualitatively by Witten {\it et al.} \cite{Witten:Pincus}.
For $f{>\atop\sim}60$, a solid-solid phase transformation into a bco phase 
takes place. This unusual phase 
is stable up to $\eta\simeq 1.0$. For $44{<\atop\sim}f{<\atop\sim}60$,
the remolten liquid refreezes into this bco structure at
$\eta\simeq0.80$. At $\eta\simeq 1.0$, a further solid-solid phase
transition from the bco into a diamond structure is found, the 
latter being stable for arm numbers $f{>\atop\sim} 44$ and packing fractions 
up to $\eta\simeq 1.4$-1.5. 
Notice that the extension of the two phase regions (``density-jumps'') of all
encountered phase transitions is extremely small due to the soft character of
$V(r)$ \cite{hartmut:hartmut}. Moreover, the empirical 
Hansen-Verlet freezing rule \cite{Hansen:Verlet} is valid for 
{\em all} points at the phase 
boundaries where we calculated the static structure factor $S(q)$. 
This also includes the reentrant melting transition for $\eta\simeq 0.7$, 
where the $S(q)$ for the fluid begins to show unusual behavior 
\cite{Watzlawek:98}.   

\begin{figure}
   {\large (a)}
   \begin{minipage}{6cm}
      \epsfxsize=5cm
      \epsfysize=4cm
      ~\hfill\epsfbox{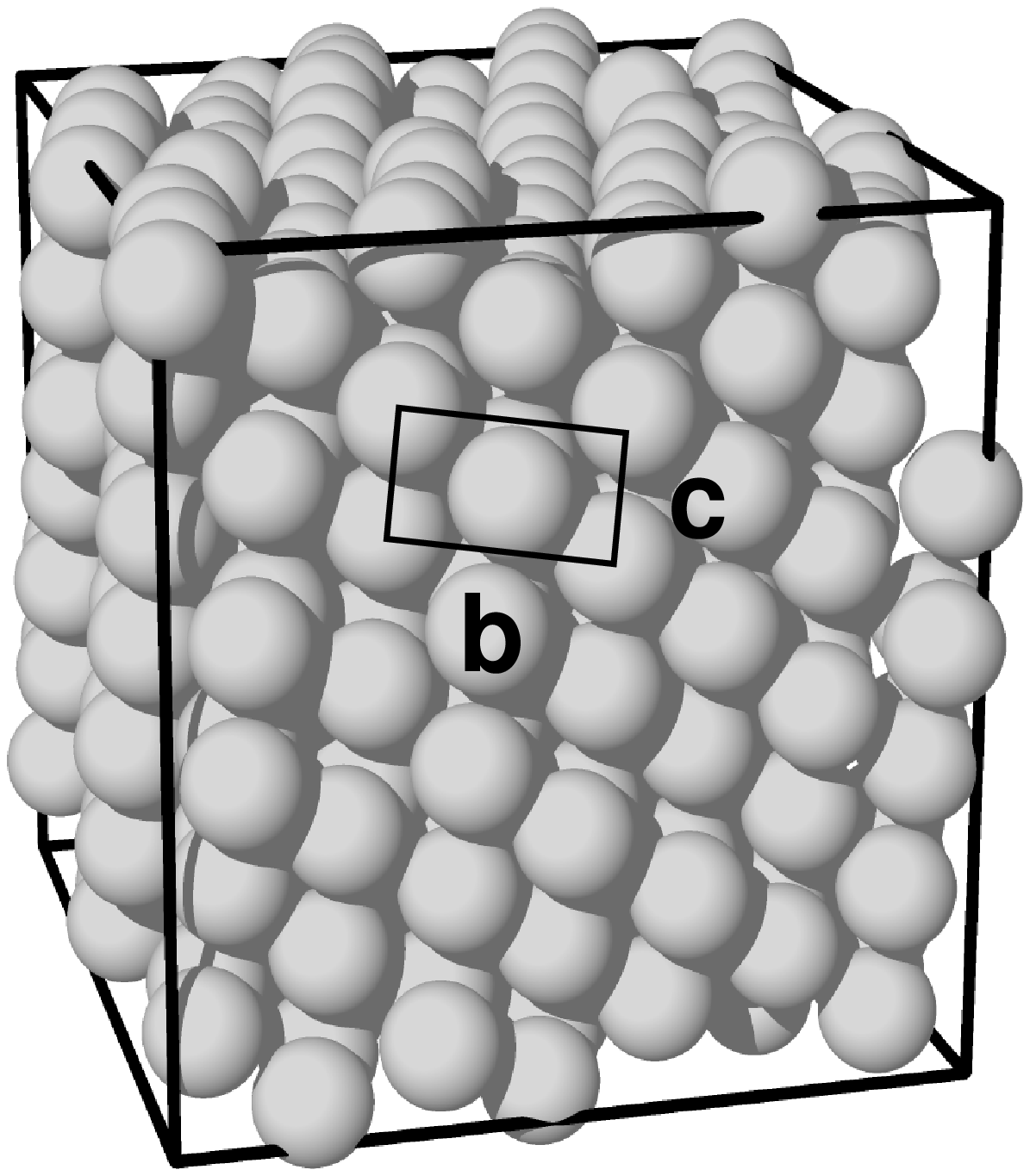}\hfill~
   \end{minipage}

   {\large (b)}
   \begin{minipage}{6cm}
      \epsfxsize=5cm
      \epsfysize=4cm
      ~\hfill\epsfbox{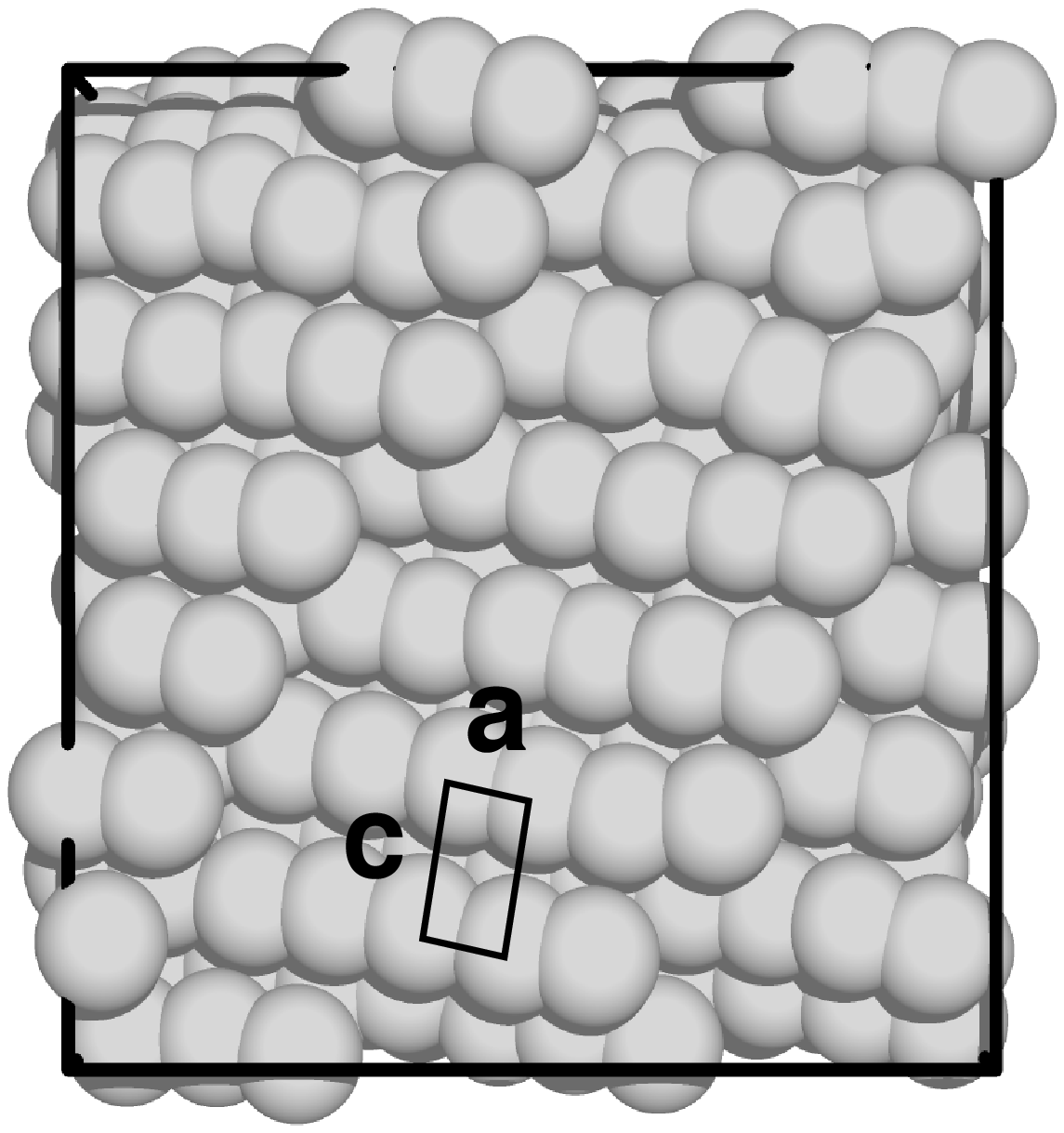}\hfill~
   \end{minipage}
  
 \caption{(a) A snapshot of a typical bco configuration for
          $f=64$ and $\eta=0.8$ in a periodically repeated (cubic) simulation
          box. The diameter of the spheres is 
          the corona diameter $\sigma$; (b) Same as (a), 
          now seen from the 
          `left' side of the simulation box shown in (a). Notice the
          high anisotropy of the lattice spacings. The elementary cell length
          ratios are $b/a \simeq 2.70$ and $c/a \simeq 1.57$.}
   \label{snapshot.plot}
\end{figure}

We develop now a physical intuition for the unusual occurrence of the 
bco and diamond phase. For this purpose, we 
report first on the detailed structure of the 
bco phase. At fixed $\eta$, the bco crystal is described by the two 
lengths ratios of its elementary cell, $b/a$ and $c/a$, respectively. 
In order to calculate the free energy of the bco crystal by the 
Frenkel-Ladd method, these ratios had to be determined from a first set of 
simulations. In these $NpT$-simulations \cite{Frenkel:Smit},
the system was free to adopt its optimal values for $b/a$ and $c/a$,
starting either from a purely random configuration or a initial bco 
configuration. Within the error bars, the so determined elementary cell 
length ratios were in agreement with the values obtained from the 
minimization of the lattice sums. We therefore took the lattice sum 
results as input for the free energy calculations. These ratios
increase with $\eta$ from $b/a\simeq 2.24$ and $c/a\simeq 1.32$ at 
$\eta=0.7$ to $b/a\simeq 3.14$ and $c/a\simeq 1.81$ at $\eta=1.0$ and 
are nearly independent of $f$. Fig.\ \ref{snapshot.plot} illustrates 
the resulting structure.
As can be seen from this figure, the anisotropy of the elementary cell
leads to a strong interpenetration of the particle coronas along one of the
lattice axes. In fact, over the whole stability range of the bco phase, 
the next neighbor distance along this axis is considerably
smaller than $\sigma$, whereas all other next neighbor
distances are larger than $\sigma$. 
This can be intuitively understood from the form of the
potential (\ref{potential}): due to the weak divergence for small $r$, there
is no huge energy penalty in bringing the nearest neighbors close together.
On the other hand, the potential falls off rapidly for $r>\sigma$, so
all the remaining neighbor shells are not costly in energy, too.
With increasing $\eta$, the distance of the two nearest neighbors in the bco
is decreasing until the energy penalty becomes significant. Hence, the bco
will then lose against another structure with more than two nearest neighbors
inside the corona. A suitable structure is the diamond phase which
possesses four tetrahedrally ordered nearest neighbors. Indeed, our simulations
show that all the other neighbors are kept outside the corona in the stability
range range of the diamond. Therefore, both the ultrasoft logarithmic part
{\em and} the crossover at $r=\sigma$ are crucial for the stability
of the bco and the diamond phase. This provides a simple reason why such phases
have not been found earlier for strongly repulsive interactions.
We further note that the presented scenario 
also nicely expresses itself in the angle-average radial distribution 
functions $g(r)$ of the bco and diamond solid, which show a similar
anomaly as found in the $g(r)$ of the fluid phase \cite{Watzlawek:98}.  

As for a further theoretical investigation, we solved the accurate 
Rogers-Young closure \cite{ry} to obtain the free energy of the fluid for
$f = 18, 32, 40, 48$ and $64$. For the aforementioned solid structures,
we used the Einstein-crystal perturbation theory \cite{einstein:theory}
to calculate the associated free energies. 
As this theory
provides only an {\it upper bound} to the free energy, the domain
of stability of the fluid is enhanced in comparison to the simulation results.
The theory predicts $40 < f_c < 48$ and eliminates the 
domain of stability of the bcc crystal. 
Otherwise, as also shown in Fig. \ref{diagram.plot}, the same phase behavior 
as determined from simulations emerges. 

We finally note that all our predictions for $\rho\le2\rho^{*}$, i.e. 
$\eta{<\atop\sim}1.0$, should be verifiable in scattering experiments,
since for these densities pair interactions are dominant.
In fact, in recent experimental work on spherical diblock copolymer
micelles, Gast and coworkers have already confirmed a part of our
results \cite{McConnell,McConnell2}.
The freezing transition in fcc and bcc crystals depending on the number of 
arms $f$ is found \cite{McConnell} as well as reentrant melting 
with increasing $\eta$ \cite{McConnell2}. For the ''most starlike''
system, also a reentrant freezing is
observed as predicted in Fig. \ref{diagram.plot}.
For $\eta{>\atop\sim}1.0$ however, when three stars exhibit overlaps within
their coronae, many body interactions become important, which we have 
neglected in our calculations using the pair potential (\ref{potential}). 
Nevertheless, from a theoretical point of view, this
potential turned out to be interesting also for 
$\eta{>\atop\sim}1.0$, resulting for the first time in a stable 
diamond structure for a purely radially symmetric pair interaction.

In conclusion, we have determined the phase diagram of star polymers over a
broad range of arm numbers $f$ and packing fractions $\eta$ by computer
simulations and theory.
The phase diagram includes a fluid phase as well as four stable
crystal phases. These crystal phases are a fcc crystal and a bcc crystal, 
as well as an unusual anisotropic bco structure and a diamond crystal.

It is a pleasure to thank Prof.\ Daan Frenkel for helpful remarks.
We further thank the Deutsche Forschungsgemeinschaft for support within
SFB 237.

\bibliographystyle{unsrt}

%
%

% 
% \section*{Figures}
% 
% \begin{figure}
%    \caption{The phase diagram of star polymer solutions for different arm 
%            numbers $f$ versus packing fraction $\eta$. 
%            The squares and the circles
%            indicate the phase boundaries as obtained from computer
%            simulations and theory, respectively; lines are only 
%            guides to the eye. The statistical error of the simulations
%            is of the order of the symbol size. The triangles indicate 
%            the freezing and melting point of hard spheres.
%           }
%   \label{diagram.plot}
% \end{figure}
% 
% \begin{figure}
%    \caption{(a) A snapshot of a typical bco configuration for
%          $f=64$ and $\eta=0.8$ in a periodically repeated (cubic) simulation
%          box. The diameter of the spheres is 
%          the corona diameter $\sigma$; (b) Same as (a), 
%          now seen from the 
%          `left' side of the simulation box shown in (a). Notice the
%          high anisotropy of the lattice spacings. The elementary cell length
%          ratios are $b/a \simeq 2.70$ and $c/a \simeq 1.57$.}
%    \label{snapshot.plot}
% \end{figure}

\end{document}